\documentclass[a4paper,aps,prl,preprintnumbers,amsmath,amssymb,twocolumn,longbibliography,superscriptaddress]{revtex4}

\usepackage{graphicx}
\usepackage{fp}
\newcount\boxheight
\newcount\boxwidth
\newcommand\testaspect[1]{%
  \setbox0=\hbox{#1}%
  \boxheight=\ht0\relax%
  \boxwidth=\wd0\relax%
  \FPdiv\theaspect{\the\boxheight}{\the\boxwidth}%
  \copy0%
}

\usepackage[dvipsnames]{xcolor}
\usepackage{graphicx}
\usepackage{dcolumn}
\usepackage{bm}
\usepackage{soul}

\usepackage[colorlinks=true,citecolor=blue,linkcolor=blue,urlcolor=blue]{hyperref}

\begin{document}

\title{Detecting Majorana Zero Modes via Strong Field Dynamics}

\begin{abstract}
    We propose a protocol to detect topological phase transitions of one-dimensional $p$-wave superconductors from their harmonic emission spectrum in strong fields. Specifically, we identify spectral features due to radiating edge modes, which characterize the spectrum and the density of states in the topological phase, and which are absent in the trivial phase. These features allow us to define an order parameter, obtained from emission measurements, that unambiguously differentiates between the two phases. 
    Local probing provides insight to the localized and topologically protected nature of the modes.The presented results establish that high harmonic spectroscopy can be used as a novel all-optical tool for the detection of Majorana zero modes.
\end{abstract}

\author{Niccol\`o Baldelli}
\affiliation{ICFO-Institut de Ciencies Fotoniques, The Barcelona Institute of Science and Technology, 08860 Castelldefels (Barcelona), Spain}
\author{Utso Bhattacharya}
\affiliation{ICFO-Institut de Ciencies Fotoniques, The Barcelona Institute of Science and Technology, 08860 Castelldefels (Barcelona), Spain}
\author{Daniel Gonz\'alez-Cuadra}%
\affiliation{ICFO-Institut de Ciencies Fotoniques, The Barcelona Institute of Science and Technology, 08860 Castelldefels (Barcelona), Spain}
\affiliation{Center for Quantum Physics, University of Innsbruck, 6020 Innsbruck, Austria}
\affiliation{Institute for Quantum Optics and Quantum Information of the Austrian Academy of Sciences, 6020 Innsbruck, Austria}
\author{Maciej Lewenstein}
\affiliation{ICFO-Institut de Ciencies Fotoniques, The Barcelona Institute of Science and Technology, 08860 Castelldefels (Barcelona), Spain}
\affiliation{ICREA, Pg. Lluis Companys 23, 08010 Barcelona, Spain}
\author{Tobias Gra{\ss}}
\affiliation{ICFO-Institut de Ciencies Fotoniques, The Barcelona Institute of Science and Technology, 08860 Castelldefels (Barcelona), Spain}
\date{\today}

\maketitle

\paragraph{Introduction --}
The past decade has witnessed a hunt for the elusive Majorana fermions (MFs)
\cite{Majorana2008}.
Although it is well established that MFs can emerge as quasiparticles from condensed matter \cite{Wilczek2009}, clear experimental evidence is still lacking. A paradigmatic system expected to host MFs at the edges is the one-dimensional (1D) spinless $p$-wave superconductor, also called Kitaev chain \cite{Kitaev_2001}. In the topological superconducting state, the MFs appear as zero-energy modes in the middle of the superconducting gap, and are therefore also called  Majorana zero modes (MZMs). They are examples of symmetry-protected topological edge states~\cite{Chiu_2016}, and in two dimensions, they possess non-Abelian anyonic statistics~\cite{Stern2010}, making them very interesting candidates for topological qubits, thanks to their capability of robustly storing and processing quantum information \cite{KITAEV20032, Sarma2015, Alicea2011}.

Despite the design of several experimental setups that effectively realize the Kitaev chain model \cite{fu08, fu09, oreg10, lutchyn10, cook11}, the detection of MZMs remains challenging. In nanowire setups, MZMs are expected to appear as zero-energy states in the tunnelling density-of-states (DOS), manifested through a quantized zero-bias peak of height $2e^2/h$ in the differential conductance \cite{law09, flensberg10, stanescu11}. Despite several experiments showing compatible results \cite{mourik12,Deng2012,Das2012}, there is still no conclusive evidence of the predicted robust quantization of the conductance \cite{Yu2021}. Specifically, the zero bias peaks are found at heights significantly smaller than $2e^2/h$, challenging their interpretation. Moreover, the observed nearly perfect conductance quantization may also stem from non-Majorana (non-topological) states \cite{liu12, lee12, dassarma2021}.

In this Letter, we develop an alternative approach to uncover MZMs 
by the non-linear response to strong sub-THz electromagnetic fields \cite{hafez18}. The strong fields bring the electrons into a non-perturbative regime in which their dynamics give rise to high-harmonic generation (HHG) \cite{vampa14}. In the past, HHG has been used to track the dynamics of excitations at femtosecond timescales, yielding ultrafast imaging methods in atomic and molecular gases \cite{krause92, krausz09, baker06, Shafir2012} and, more recently, in solid-state systems \cite{Hohenleutner2015, Ghimire2019, alcala22}. Lately, there has been a rising interest in using HHG to detect topological properties of matter \cite{reimann2018subcycle, bauer18, bauer19, Silva2019, chacon2020,pattanayak21,shao22}.

Here, we analyze the HHG from a proximity-induced 1D $p$-wave superconductor, showing that it directly reflects the density of states, and thus, the width of the energy bands and energy gaps. This gives rise to a spectroscopic scheme that distinguishes the topological from the trivial phase. Specifically, we introduce an order parameter, obtained from the HHG spectrum, to chart out the whole phase diagram. Moreover, by focusing the radiation source to the edge, we are able to discern bulk from edge excitations, clearly isolating the contribution from the topological MZMs. 
Our method can be used as an independent check for the presence of MZMs, which, as an all-optical technique, can easily be applied to any sample without the need for gating, and which provides unprecedented spatial and temporal resolution. Thus, it is suited to complement or even substitute transport-based detection techniques \cite{law2009}, in order to provide the sought-after evidence of MZMs in condensed matter setups.

\paragraph{Model --}
The 1D $p$-wave superconducting Kitaev chain \cite{Kitaev_2001} is described by the Hamiltonian $H_{K}=\sum_n [-\mu c_n^{\dagger}c_n-t(c^{\dagger}_{n+1}c_n+{\rm h.c.})+\Delta(c_nc_{n+1}+{\rm h.c.})]$,
where $c^\dagger_n$ ($c_n$) are creation (annihilation) operators of spinless fermions on site $n$. This model exhibits two phases, a trivial one and a topological one, with a topological phase transition at $|\mu|=2t$. In the topological phase, for open boundary conditions, the spectrum is characterized by two degenerate ground states, corresponding to MZMs localized at the two edges of the chain.

For studying the response of this model to a strong field, the fermions' coupling to the electromagnetic field is crucial. A naive coupling to the vector potential $A(t)$ via Peierls substitution, $c_j^{\dagger}\rightarrow e^{iAj}c_j^{\dagger}$, would require a corresponding dynamical change in the superconducting gap $\Delta$ to preserve the gauge-invariance of the Hamiltonian. While this is possible, we make the approximation that the value of 
$\Delta$ remains fixed throughout. 
Therefore, we focus on a particular system \cite{oreg10}, which has been the main focus of recent experimental investigations and whose low energy behavior is governed by the Kitaev chain Hamiltonian: a heterostructure between a semiconducting chain with strong spin-orbit coupling and a regular $s$-wave superconductor, additionally subjected to an external Zeeman field. Since the interactions with a strong field also excite high energy states, we study the full multiband Hamiltonian of the heterostructure and not just its low energy subspace. To capture the overall dynamics, we consider the time-dependent Hamiltonian for a chain of $N$ sites in the Bogoliubov-de Gennes basis,
$H=\Psi^{\dagger}H_{BdG}\Psi$, with
\begin{equation}
H_{BdG} (t)=
\begin{pmatrix}
\setlength{\arraycolsep}{0pt}
\renewcommand{\arraystretch}{0.8}
J+I\frac{B}{2}&U(t)&I\Delta&0\\
U^{\dagger}(t)&J-I\frac{B}{2}&0&I\Delta\\
I\Delta^*&0&-J+I\frac{B}{2}&-U(t)\\
0&I\Delta^*&-U^*(t)&-J-I\frac{B}{2}\\
\end{pmatrix}.
\end{equation}
Here, $I$ is the $N \times N$ identity matrix, and $J$ and $U$ are $N\times N$ matrices defined by
$ J_{l,m} = -\mu \delta_{l,m} + (\delta_{l,m-1} j e^{iA(t)} +{\rm H.c.})$ and $ U_{l,m}(t) =  \delta_{l,m-1} \alpha e^{iA(t)} +{\rm H.c.}$. The operator $\Psi^{\dagger}=(c^{\dagger}_{\uparrow},c^{\dagger}_{\downarrow},c_{\downarrow},-c_{\uparrow})$ is a ``compressed" Nambu spinor, where $c_\sigma\equiv c_{1\sigma},\dots,c_{N\sigma}$, with $\sigma \in \{\uparrow, \, \downarrow\}$. The Hamiltonian's parameters are the hopping $j$, the chemical potential $\mu$, the effective spin-orbit coupling $\alpha$, the Zeeman field $B$ and the proximity-induced superconducting coupling $\Delta$. 
The structure of the Hamiltonian allows us to straightforwardly couple the system to the external field via a Peierls substitution. For convenience, we choose units of $A(t)$ such that the coupling constant $e a/\hbar$, with $a$ the unit cell size, is 1. The time-dependence of the vector potential is of the form
\begin{equation}
        A(t)=A_{0}\sin(\omega t)\sin^2\left(\frac{\omega t}{2 n_c}\right)-\varepsilon_c t, \quad 0\leq t\leq 2\pi n_c/\omega
\end{equation}
describing a pulse of $n_c$ cycles with frequency $\omega$, and a constant electric field with $\varepsilon_c$, explicitly breaking time inversion symmetry in the system, such that both even and odd harmonics of the driving frequency can be generated \cite{kanega21}.
Measuring energies in units of $j$, we choose $\omega=0.0025$, such that it corresponds roughly to $1/50$ of the bandgap of the system. The symmetry-breaking DC field is very weak, $\varepsilon_c=10^{-5}$, whereas the amplitude of the vector potential has to be strong enough to produce high harmonics and is taken to be $A_{0}=1.2$, which for $a=0.5$ nm corresponds to $1.6\times 10^{-6}$ Vs/m. 

For the static Hamiltonian at $t=0$, the topological phase appears for $B>\sqrt{\Delta^2+\mu^2}$~\cite{leijnse2012}. For lower values of $B$, the system is in a trivial gapped superconducting state with no topological edge modes.  We refer to the Supplemental Material for a full derivation of the Hamiltonian. Although in realistic semiconductor/superconductor heterostructures, the energy scales $j,\,\alpha,\,\Delta$ widely differ, in the following we choose them to be of the same order of magnitude (specifically, $\alpha=\Delta=3/4$). The reason for this choice is to achieve clear Majorana modes for system sizes that are sufficiently small to numerically perform simulations of the full dynamics. In particular, the parameters were tuned according to the prescription from \cite{alicea2012} in order to assure the presence of a topological phase. 
For the purpose of detecting MZM from the high harmonic spectrum, it is important that the ratio between the bandgap (usually of the order of $\Delta$) and the frequency $\omega$ of the incoming pulse is much larger than 1. For InAs nanowires, the bandgap is on the order of 1 meV (with $j\sim3000$ meV, $\alpha\sim 25$ meV, $\Delta\sim B\sim\mu\sim1$ meV, cf. Ref.~\onlinecite{stanescu11}), but also much larger gaps have been reported, of 4 meV for $\beta$-Bi$_2$Pd films \cite{guan19}, or even 15 meV for iron-based superconductors \cite{hagiwara21}. Depending on the size of the gap, our scheme requires strong microwave to THz sources \cite{hafez18}, with pulse duration on the order of 1-100 ps, which is potentially much shorter than typical relaxation time scales.




\begin{figure}
\centering
\includegraphics[width=\linewidth]{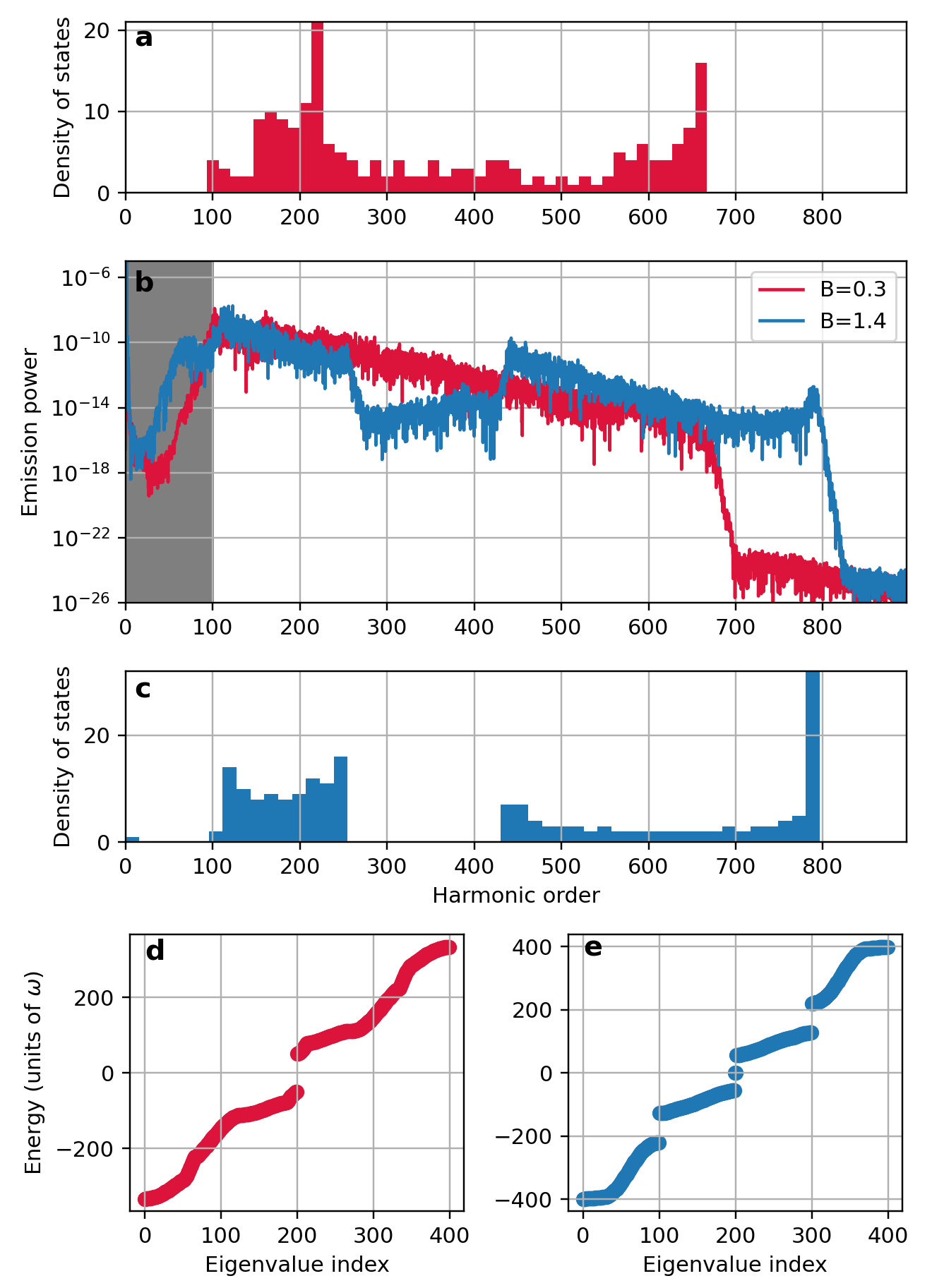}
    \caption{Comparison of the emitted spectrum deep in the topological ($B = 1.4$) and trivial phases ($B = 0.3$) for a spatially uniform field, and $\mu=0$. The emission spectra shown in (b) follow the density of state of the Hamiltonian at time $t=0$ both in (a) the trivial and (c) topological phases. For the topological phase, an emission below the bandgap of the system (gray are in (b)) can be related to the presence of zero energy edge states. The density of states can be compared with the band structure in the trivial (d) and topological (e) phases.}
    \label{fig:spectrum}
\end{figure}

\paragraph{High-harmonic generation --} The key quantity that captures the non-linear optical response of the system is the transmitted HHG spectrum, that is, the power spectrum of emission: 
\begin{equation}
    P(\omega)\propto|\text{FT}(\dot{\langle x\rangle})|^2,
    \label{fft}
\end{equation}
normalized to the maximum of the power spectrum of the incoming pulse \cite{baggesen:hal-00625578}. Here, the time derivative of the average dipole moment $e\langle x(t)\rangle$ yields the electric current, which is Fourier transformed into the frequency domain. The dipole moment is calculated by numerically integrating the time-dependent Schr\"odinger equation (TDSE) from the initial ground state of the Hamiltonian up to a time $T=2\pi n_c/\omega$ (see Supplemental Material for details).

In Fig.~\ref{fig:spectrum}(b) we plot the emission spectrum, obtained from Eq.~\eqref{fft}, as a function of the harmonic order of the driving frequency. We consider two points in parameter space, one deep in the trivial phase at $B=0.3$ (red), and the other one in the topological one at $B=1.4$ (blue). Interestingly, we observe that, in both cases, the spectrum echoes the band structure of the Bogoliubov Hamiltonian, plotted in Fig.~\ref{fig:spectrum}(d) and (e), with two or four bands symmetric around the Fermi energy due to particle-hole symmetry \cite{alicea2012}. In particular, the density of states in the two phases, shown in Fig.~\ref{fig:spectrum}(a) and (c), clearly determines the emission. In particular, there is no emission above the bandwidth and below the bandgap (defined as the difference between the highest valence band and the lowest conduction band excluding edge modes) in the trivial phase. In stark contrast, in the topological phase, the radiation plateau starts from half-bandgap, which is related to the presence of radiating edge modes at zero-energy.

For a qualitative understanding of the emission spectrum, we note that the dynamics that cause the emission can be split into three different steps \cite{vampa14}: 1) the incoming pulse can excite a Bogoliubov quasiparticle from the filled bands to the empty ones; 2) the excitation can then move inside the empty band under the applied electric field, and, subsequently, 3) it can relax back to one of the occupied bands.  
This leads to two different kinds of contributions in the emission spectrum, an intraband one (step 2) and an interband one (steps 1 and 3). The intraband contribution is 
produced by the acceleration (Bloch oscillations) of the quasiparticles within a  
band with non-linear dispersion. The frequency of the interband emission, on the other hand,
is bounded by the bandgap (lowest possible interband excitation) and the bandwidth of the system (highest one)~\footnote{We note that, for a short envelope as considered here ($n_c=5$), the Fourier transform of the incoming pulse is spread around the driving frequency. There is subsequently a widening and possible mixing of the harmonics generated by the system, but this is not of particular concern for our purposes as we are not interested in discerning the emission at a particular frequency.}.

\begin{figure}
    \centering
    \includegraphics[width=\linewidth]{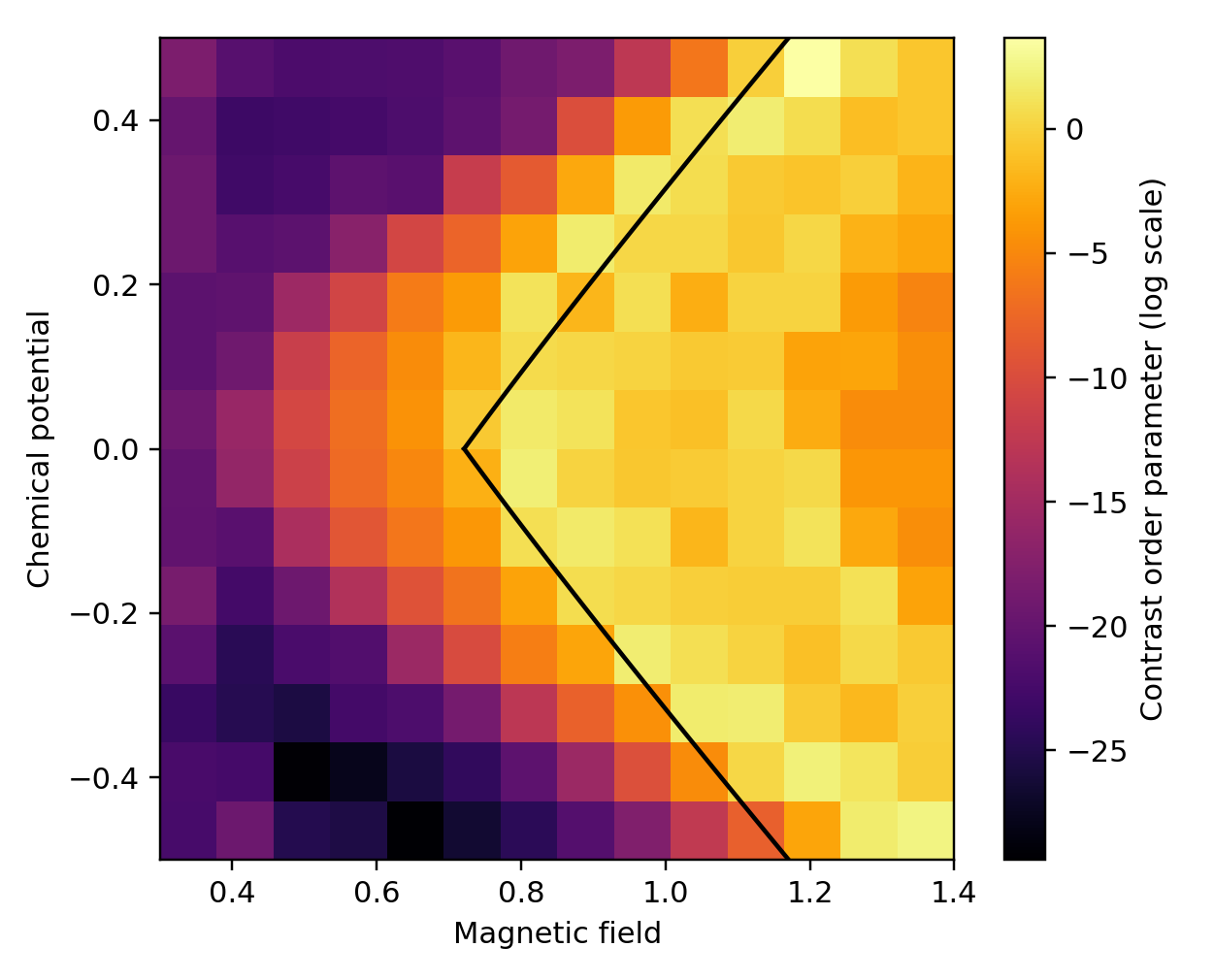}
    \caption{Contrast order parameter $C$ as a function of magnetic field and chemical potential. In the topological phase $C\sim1$, while in the trivial one $C\sim 0$. The black line represents the phase separation boundary for a system in the thermodynamic limit. 
 }
    \label{fig:phased}
\end{figure}

\paragraph{Order parameter --}
We now propose a contrast order parameter, defined as the ratio between the emission at half the bandgap $P_{\rm half}$ over the emission at the bandgap $P_{gap}$,
\begin{equation}
    C=\frac{\log(P_{\rm half})}{\log(P_{\rm gap})},
\end{equation}
which is of order one in the topological phase and zero in the trivial phase. The topological phase diagram of the system is computed  Fig.~\ref{fig:phased} in the $\mu-B$ plane using the proposed order parameter. Exactly at the boundary where the gap closes, the system behaves as a metal, and our order parameter is greater than one here as the emission is higher for lower harmonics. 
The choice of the frequency of the incident light pulse is crucial to  localize the phase boundary, as a lower frequency provides a sharper criterion for distinguishing topological and trivial phases. From this point of view, choosing a small driving frequency is favorable, as long as the pulse remains short as compared to relaxation times. We refer to the Supplemental Material for a discussion of this behavior.




\begin{figure}
    \centering
    \includegraphics[width=\linewidth]{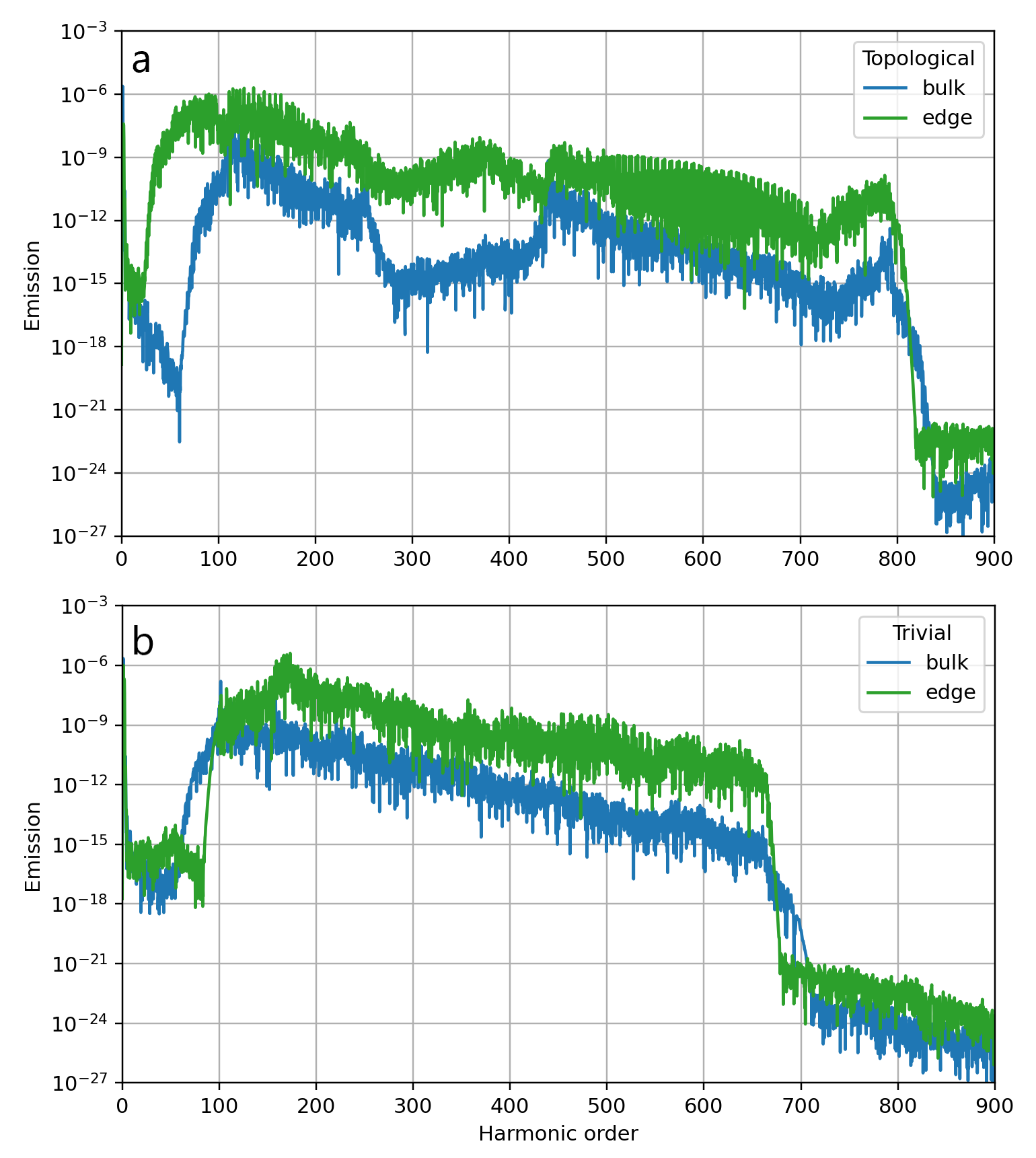}
    \caption{Emission for a pulse focused on the edge or on the bulk of the sample in the topological (a) and trivial (b) phases. The spacial envelope is normalized to have constant energy of the EM field. In the topological phase, the spectra differ depending on where the pulse is focused, showing emission from sub band-gap states only when it is focused on the edge. This is not the case, however, in the trivial phase, where both spectra are qualitatively similar.}
    \label{fig:edgebulk}
\end{figure}

\paragraph{Robust edge states --}
So far, the devised scheme distinguishes between topological and trivial phases by measuring the full bandstructure of the system, but yet it does not capture the maybe most stunning property of the MZMs, their localization at the edge and topological protection. However, with the spatial resolution of the radiation being limited only by the wavelength, it becomes possible to demonstrate that the sub-bandgap emission is due to edge modes by focusing the electromagnetic field either on the edge or the bulk of the sample. In Fig.~\ref{fig:edgebulk}, we show how in the trivial phase the emitted spectrum is qualitatively the same for a pulse focused on the edge or on the bulk. On the other hand, in the topological phase, there is strong radiation between the bandgap and half bandgap if the light is focused on the edge, showing that the contribution to the emitted spectrum in this mid-band gap region does not come from the bulk, but solely from the edge. We have used a Gaussian envelope, and the amplitude of the envelopes is normalized in order to have the same electromagnetic energy for all cases (edge, bulk, and uniform field).

\begin{figure}
    \centering
    \includegraphics[width=\linewidth]{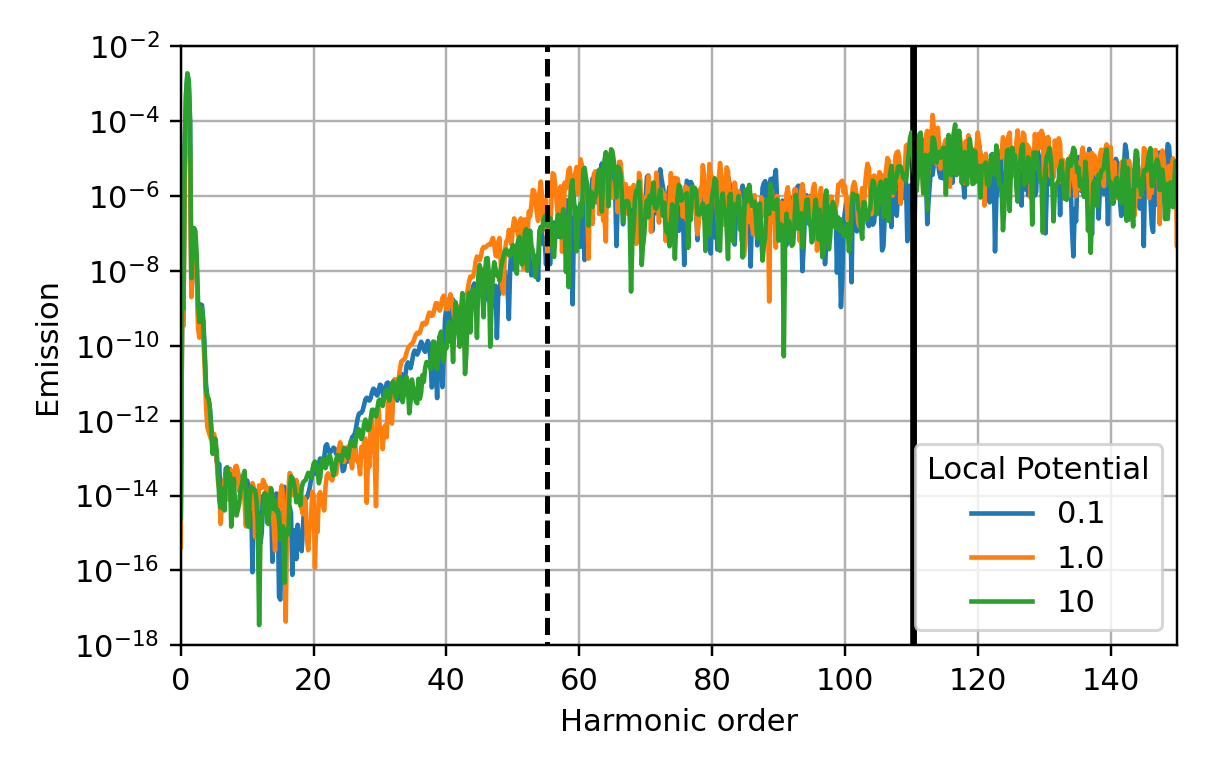}
    \caption{Emission spectra with the addition of a local potential on the edges of the chain (in units of $j$). The local potential acts on the first and last 3 sites and does not break the symmetry responsible for the topological protection. The qualitative behavior of the emission of the edge states does not change with an increasing value of this potential. The black lines indicate half bandgap and the bandgap in order of the driving frequency.}
    \label{fig:phdcontrast}
\end{figure}

Finally, to illustrate the topological nature of the edge modes, we add a local perturbation to the system that does not break the particle-hole symmetry of the Hamiltonian. Such local potentials (acting on three sites on the left and right edges) can be added by applying gate voltages at the edges and are modeled by adding a term $j\ell_p\sum_{i,\sigma}c^{\dagger}_{i\sigma}c_{i\sigma}$ for $i=1,2,3,N-2,N-1,N$. We observe that increasing the value of this potential does not affect the shape of the emission spectrum. Specifically, the sub-bandgap edge state emission only appears in the topological phase and remains present precisely between half bandgap and the bandgap, as can be seen in Fig.~\ref{fig:phdcontrast}. The different curves in this figure, corresponding to different edge potentials, lie on top of each other for lower harmonics (until the bandgap of the system), indicating that the edge modes do not shift in energy upon the application of edge potentials, clearly demonstrating the topological robustness of the radiating edge modes.

\paragraph{Summary and outlook --} 
In the present Letter, topological edge modes are detected via the electromagnetic emission spectrum in the non-linear regime. An experimentally observable quantity, the contrast order parameter, is constructed to map the phase diagram. To confirm the topological nature of radiating modes, the system can be probed locally and shown to be robust under local perturbations. Thus, our protocol complements other established methods in the pursuit for MZMs.

A major experimental challenge for the  detection of MZMs is to distinguish them from trivial sub-bandgap states that can appear in the material, that is, Andreev bound states \cite{sauls2018andreev, Prada2020}. These states can appear from regions in the semiconducting chain where the proximity-induced superconductivity fails (i.e. near the edges), creating zones of normal metal where scattering effects can lead to the creation of localized states \cite{andreev1964thermal}. A straightforward extension of our work will use a spatially dependent superconducting order parameter to study the formation of these states and their influence on the emission spectrum. Another interesting application of our technique is the study of Majorana physics in two dimensions, where MZMs can arise as vortices in $p$-wave superconductors \cite{read00}. Compared to 1D models, the dependence of  the emission spectrum on the field polarization states will make the two-dimensional scenario very rich: Recent studies have shown that left-right polarized drives can shed light on the presence of topological chiral states \cite{tran17, Asteria2019}, both in the perturbative  and ultra-strong regime \cite{Silva2019, chacon2020}, but a study of the effect on Majoranas is still missing.

\begin{acknowledgments}
We thank A. Dauphin, A. Maxwell, J.-M. Raimond, and P. Stammer  for useful discussions and insightful comments. ICFO group acknowledges support from: ERC AdG NOQIA; Agencia Estatal de Investigaci\'on (R\&D project CEX2019-000910-S, funded by MCIN/ AEI/10.13039/501100011033, Plan National FIDEUA PID2019-106901GB-I00, FPI, QUANTERA MAQS PCI2019-111828-2, Proyectos de I+D+I ``Retos Colaboraci\'on" RTC2019-007196-7); Fundaci\'o Cellex; Fundaci\'o Mir-Puig; Generalitat de Catalunya through the CERCA program, AGAUR Grant No. 2017 SGR 134, QuantumCAT/U16-011424, co-funded by ERDF Operational Program of Catalonia 2014-2020; EU Horizon 2020 FET-OPEN OPTOLogic (Grant No 899794); National Science Centre, Poland (Symfonia Grant No. 2016/20/W/ST4/00314); Marie Sk\l odowska-Curie grant STREDCH No 101029393; ``La Caixa" Junior Leaders fellowships (ID100010434) and EU Horizon 2020 under Marie Sk\l odowska-Curie grant agreement No 847648 (LCF/BQ/PI19/11690013, LCF/BQ/PI20/11760031, LCF/BQ/PR20/11770012, LCF/BQ/PR21/11840013). N.B. acknowledges support from a ``la Caixa" Foundation (ID 100010434) fellowship. The fellowship code is  LCF/BQ/DI20/11780033.
T.G. acknowledges financial support from a fellowship granted by ``la Caixa” Foundation (ID 100010434, fellowship code LCF/BQ/PI19/11690013). D.G.-C. is supported by the Simons Collaboration on Ultra-Quantum Matter, which is a grant from the Simons Foundation (651440, P.Z.). 
\end{acknowledgments}


\bibliography{final}

\end{document}


\onecolumngrid
\section{\label{SM} Supplemental material}

\subsection{Hamiltonian of the system}

We start from the continuum Hamiltonian of the semiconducting chain:

\begin{equation}
    H_0=\sum_{s}\int\dif x \psi_s^{\dagger}(x)\left(\frac{p^2}{2m}-\mu+V(x)-\alpha E_{\perp}p\sigma_y +\frac{1}{2}g\mu_B B\sigma_z\right)\psi_s(x)
\end{equation}

Here $\sigma$ are Pauli matrices in spin space, $s$ is a spin index, $B$ is a Zeeman field in the $z$ direction and $\alpha E_{\perp}$ is the spin-orbit coupling. $V(x)$ is the lattice potential that the electrons are subjected to.

On the other hand we can model the proximity-induced $s$-wave  superconducting coupling as:

\begin{equation}
    H_{sc}=\int\dif x\psi_{\downarrow}(x)\Delta\psi_{\uparrow}(x)+h.c.
\end{equation}
where we supposed coupling between electrons in the same position.

We can write the the coupling to the electromagnetic field of the external pulse $A(x,t)$ (that we assumed to point in the $x$ direction) as:

\begin{equation}
    H_0=\sum_{s}\int\dif x \psi_s^{\dagger}(x)\left(\frac{(p-eA)^2}{2m}-\mu+e\Phi+V(x)-\alpha E_{\perp}(p-eA)\sigma_y +\frac{1}{2}B\sigma_z\right)\psi_s(x),
\end{equation}

We then switch to a Wannier function basis, while applying a gauge transformation to add a phase term to simplify the computation of the terms involving the field $A$.
\begin{equation}
    \psi_s^{\dagger}(x)=\sum_n e^{-i\lambda_n(x,t)}\phi_n^*(x)c_{ns}^{\dagger}(t), \qquad \lambda_{n}=e\int_{X_n}^{x}\dif x A(x,t),
\end{equation}
where $\phi_n(x)$ are generalized Wannier function centered on the lattice sites at $X_n$ and $c^{\dagger}_{ns}$ are the related creation operators.

Under the assumption that the EM field wavelength is much larger than the lattice spacing we can employ the dipole approximation and:
\begin{equation}
    A(x,t)\sim A(t), \qquad \lambda_n\sim eA(t)(x-X_n).
\end{equation}

The Hamiltonian can then be simplified as:
\begin{equation}
\begin{split}
    H_0=\sum_{m,n,s}\left[\int\dif x \phi_m^{*}(x)e^{-i(\lambda_m-\lambda_n)}\left(\frac{p^2}{2m}-\mu(t)+V(x)-\alpha E_{\perp}p\sigma_y +\frac{1}{2}B\sigma_z\right)\phi_n(x)\right]
\end{split}
\end{equation}

We now assume tight binding of the electron in the atomic position (as given by the potential $V(x)$). The Hamiltonian becomes:
\begin{equation}
\begin{split}
    H_0=\sum_{m,n,s}J_{mn}(t)c^{\dagger}_{ms}c_{ns}+\sum_{m,n,s,s'}V_{mn,s's}(t)c^{\dagger}_{ms'}c_{ns}
\end{split}
\end{equation}

where
\begin{eqnarray}
    J_{mn}(t)=Je^{-iA(t)(X_m-X_n)}[\delta_{m+1,n}+\delta_{m-1,n}]-\mu(t)\delta_{m,n}\\
    V_{mn,s's}(t)=-i\Tilde{\alpha}\sigma_{y,s's}e^{-iA(t)(X_m-X_n)}[\delta_{m+1,n}-\delta_{m-1,n}]+\frac{1}{2}B\sigma_{z,s's}\delta_{m,n}.
\end{eqnarray}

The summation in one of the site indices is trivial due to the Kronecker symbols, and one arrives at the form of the Kitaev chain Hamiltonian given in the main text. Note that the proximity-induced superconducting pairing, only containing operators acting on the same site (two of which are ``hidden" in the order parameter $\Delta$), does not couple with the electromagnetic field.

\section{Time evolution}

We now discuss the problem of solving the Time Dependent Schr\"odinger equation 
\begin{equation}
    i\partial_t\psi_i(t)=H(t)_{ij}\psi_j(t)
\end{equation}
where $H(t)$ is defined in the main text and $\psi_i(0)$ is an eigenstate of $H(0)$ (we will then consider $N$ of these eigenstates to discuss the many body problem).

The solution of this problem is generically 
\begin{equation}
     \psi_i(t)=\cal{T}\exp\left(-i\int_0^t H(t')_{ij}dt'\right)\psi_j(0)
\end{equation}
where $\cal{T}$ is the time-ordering operator. This evolution can be approximated for short times as 
\begin{equation}
    \psi_i(t+dt)=\exp(-iH(t)_{ij}dt)\psi_j(t)
\end{equation}

We then approximate the action of the exponential on the state either via a Crank-Nicholson approximation:
\begin{equation}
    \exp(-iH(t)_{ij}dt)=\frac{1-iH(t)_{ij}dt/2}{1+iH(t)_{ij}dt/2}+O(dt^3)
\end{equation}
where the term at the denominator is included to ensure the unitarity of the evolution. 
The Fourier Transform of the current is computed using the Scipy FFT function, with the use of a Hanning window to avoid spreading of low frequency contributions that would mask the higher part of the spectrum. The choice is motivated by the need to analyze the average emission over a large range of frequencies instead than resolving the single harmonics. To avoid aliasing artifacts related to the undersampling of the dynamics of the system, it is necessary to implement a time step $\delta t$ that is smaller than the inverse of the highest frequency scale of the system (the bandwidth) $\sim 2j$.

\section{Scaling with driving frequency}

\begin{figure}
    \centering
    \includegraphics[width=0.5\linewidth]{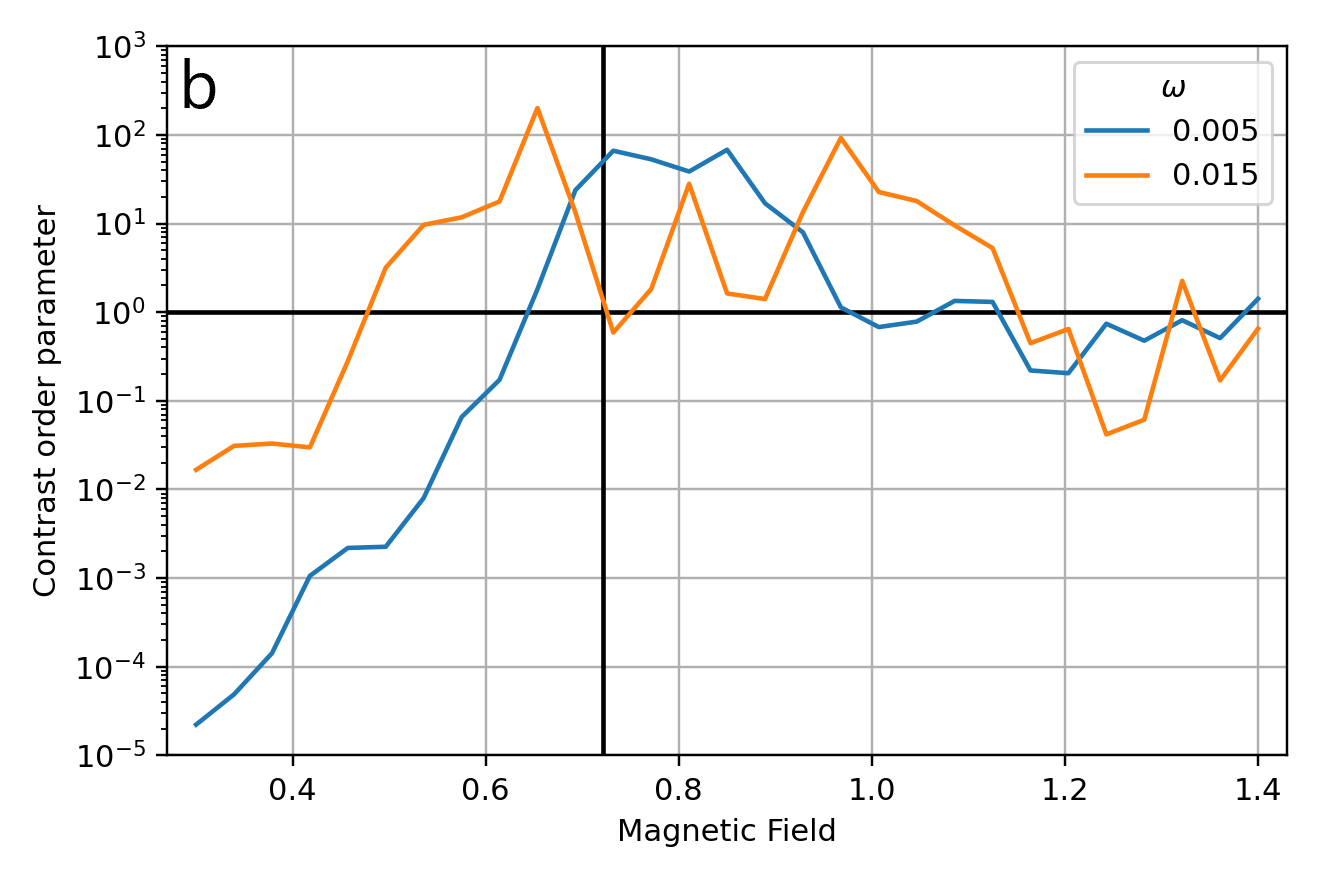}
    \caption{Contrast order parameter as a function of magnetic field (at $\mu=0$) for two values of the driving frequency. For the smaller frequency, the order parameter is peaked at the phase boundary (indicated by the vertical black line), and quickly drops below 1 (indicated by the horizontal black line) in the trivial phase. For the larger frequency, the distinction between the phases is less sharp, and the order parameter over-estimates the topological regime.}
    \label{fig:scaling}
\end{figure}

To properly explore the topological phase transition the driving frequency has to be carefully chosen: as the system approaches the transition the gap closes and when the driving frequency $\omega$ is of the order of the bandgap $\sim \Delta$ the response of the system becomes metallic. In this regime the contrast order parameter is higher than 1 and generally not stable as shown in Fig.~\ref{fig:scaling}, where the bandgap deep in the trivial phase is $\sim\Delta=0.4$. The phase transition is then better resolved the lower the driving frequency is set.

